# Stable Compound of Helium and Sodium at High Pressure

Xiao Dong,[1, 2, 3] Artem R. Oganov,[3-6*], Alexander F. Goncharov,[7,8], Elissaios Stavrou,[7,9] Sergey Lobanov,[7,10] Gabriele Saleh,[5] Guang-Rui Qian,[3] Qiang Zhu,[3], Carlo Gatti,[11] Volker L. Deringer[12], Richard Dronskowski[12], Xiang-Feng Zhou,[1, 3*] Vitali Prakapenka,[13] Zuzana Konôpková,[14] Ivan Popov,[15] Alexander I. Boldyrev[15] and Hui-Tian Wang[1,16*]

[1]School of Physics and MOE Key Laboratory of Weak-Light Nonlinear Photonics, Nankai University, Tianjin 300071, China

[2]Center for High Pressure Science and Technology Advanced Research, Beijing 100193, China

[3]Department of Geosciences, Stony Brook University, Stony Brook, New York 11794-2100, USA

[4]Skolkovo Institute of Science and Technology, 3 Nobel St., Moscow 143026, Russia

[5]Moscow Institute of Physics and Technology, 9 Institutskiy Lane, Dolgoprudny city, Moscow Region, 141700, Russia

[6]International Centre for Materials Discovery, Northwestern Polytechnical University, Xi'an, 710072, China

[7]Geophysical Laboratory, CIW, 5251 Broad Branch Road, Washington, D.C. 20015, USA

[8] Key Laboratory of Materials Physics and Center for Energy Matter in Extreme Environments, Institute of Solid State Physics, Chinese Academy of Sciences, 350 Shushanghu Road, Hefei, Anhui 230031, China

[9]Lawrence Livermore National Laboratory, Physical and Life Sciences Directorate, P.O. Box 808 L-350, Livermore, California 94550, United States

[10]Sobolev Institute of Geology and Mineralogy Siberian Branch Russian Academy of Sciences, 3 Pr. Ac. Koptyga, Novosibirsk 630090, Russia.

[11]Istituto di Scienze e Tecnologie Molecolari del CNR (CNR-ISTM) e Dipartimento di Chimica, Universita' di Milano, via Golgi 19, 20133 Milan, Italy

[12]Chair of Solid-State and Quantum Chemistry, RWTH Aachen, D-52056 Aachen, Germany

[13]Center for Advanced Radiation Sources, University of Chicago, Chicago, IL 60637, USA

[14]Photon Science DESY, D-22607 Hamburg, Germany

[15]Department of Chemistry and Biochemistry, Utah State University, Logan, Utah 84322, USA.

[16] Collaborative Innovation Center of Advanced Microstructures, Nanjing University, Nanjing, 210093, China

**Helium, on par with neon, is the most chemically inert element in the Periodic Table. Due to its extremely stable closed-shell electronic configuration with a record-high ionization potential and zero electron affinity, helium is not known to form thermodynamically stable compounds, except a few inclusion compounds. Here, using the *ab initio* evolutionary algorithm USPEX and subsequent high-pressure synthesis in a diamond anvil cell, we report the discovery of a thermodynamically stable compound of helium and sodium, $Na_2He$, which has a fluorite-type structure and is stable at pressures >113 GPa. We show that the presence of He atoms causes strong electron localization and makes this material insulating. This phase is an electride, with electron pairs localized in interstices, forming eight-centre two-electron bonds within empty $Na_8$ cubes. We also predict the existence of $Na_2HeO$ with a similar structure at pressures above 15 GPa.**

Helium (He) is the second (after hydrogen) most abundant element in the universe, and plays an enormous role in normal stars and gas giant planets, such as Jupiter and Saturn[1]. Helium and neon are the most inert elements in the Periodic Table. This is easy to understand as the ionization potential of the He atom (24.59 eV)[2] is highest among all elements, and its electron affinity is zero[3]. In the last decades, many scientists tried to find stable compounds of helium. The most successful example is the $HeH^+$ radical[4] (and, in general, $He_nH^+$ radicals, n=1-6), stable only in the charged form, extremely aggressive and protonating any base. All neutral molecules that have been found in theory or experiment, for examples, HHeF[5], (HeO)(CsF)[6], and LiHe[7], are metastable and very high in energy. For instance, HHeF has the computed energy of more than 2 eV/atom higher than the mixture of HF molecules and He atoms. The only known stable solid compounds involving helium are van der Waals compounds, such as $NeHe_2$[8] and He@$H_2O$[9]. For inclusion compounds, the enthalpy of formation is close to zero, and removal of He atoms has little effect on host's electronic structure.

Pressure greatly affects chemistry of the elements – e.g., heavy noble gases become more reactive and form compounds with both electronegative and electropositive elements, such as xenon oxides[10,11] and Mg–NG (NG = Xe, Kr, Ar)[12]. Metallic sodium, when subjected to the pressure of 200 GPa, becomes an insulator due to strong core-core orbital overlap leading to interstitial valence electron localization[13]. Furthermore, unexpected compounds, such as $Na_3Cl$, $Na_2Cl$, $Na_3Cl_2$, $NaCl_3$, and $NaCl_7$[14], become stable under pressure. Here, we performed a large-scale evolutionary search for possible stable compounds of helium with a variety of elements (H, O, F, Na, K, Mg, Li, Rb, Cs, etc.). We found that only Na readily forms a stable compound with He at pressures accessible to static experiments. Below, we focus on the Na-He system.

Searches for stable compounds were done using the variable-composition evolutionary structure prediction algorithm[15], as implemented in the USPEX code[16]. In such calculations, a phase is deemed stable if its enthalpy of formation from the elements or any other possible compounds is negative. Variable-composition structure searches were performed for the Na-He system at pressures of 0, 150, 200, 400, 700 and 1000 GPa, allowing up to 36 atoms per primitive cell. We found a new compound $Na_2He$ (Figs. 1 and 2) that has lower enthalpy than the mixture of elemental Na and He, or any other mixture, at pressures above 160 GPa (Fig. 1). The reaction

$$2Na + He \rightarrow Na_2He \quad (1)$$

is predicted to be exothermic at pressures above 160 GPa, releasing as much energy as -0.51 eV at 500 GPa. Phonon calculations clearly indicate dynamical stability of $Na_2He$ above 100 GPa (Supplementary Fig. S1). This means that, once formed, this phase can be quenchable down to 100 GPa, but at and below 50 GPa it is dynamically unstable and therefore unquenchable to ambient conditions. Quasiharmonic free energy calculations suggest that temperature has little effect on Gibbs free energy of formation of $Na_2He$: e.g., it increases from -0.41 eV at 0 K to -0.39 eV at 800 K at 300 GPa (Supplementary Fig. S2).

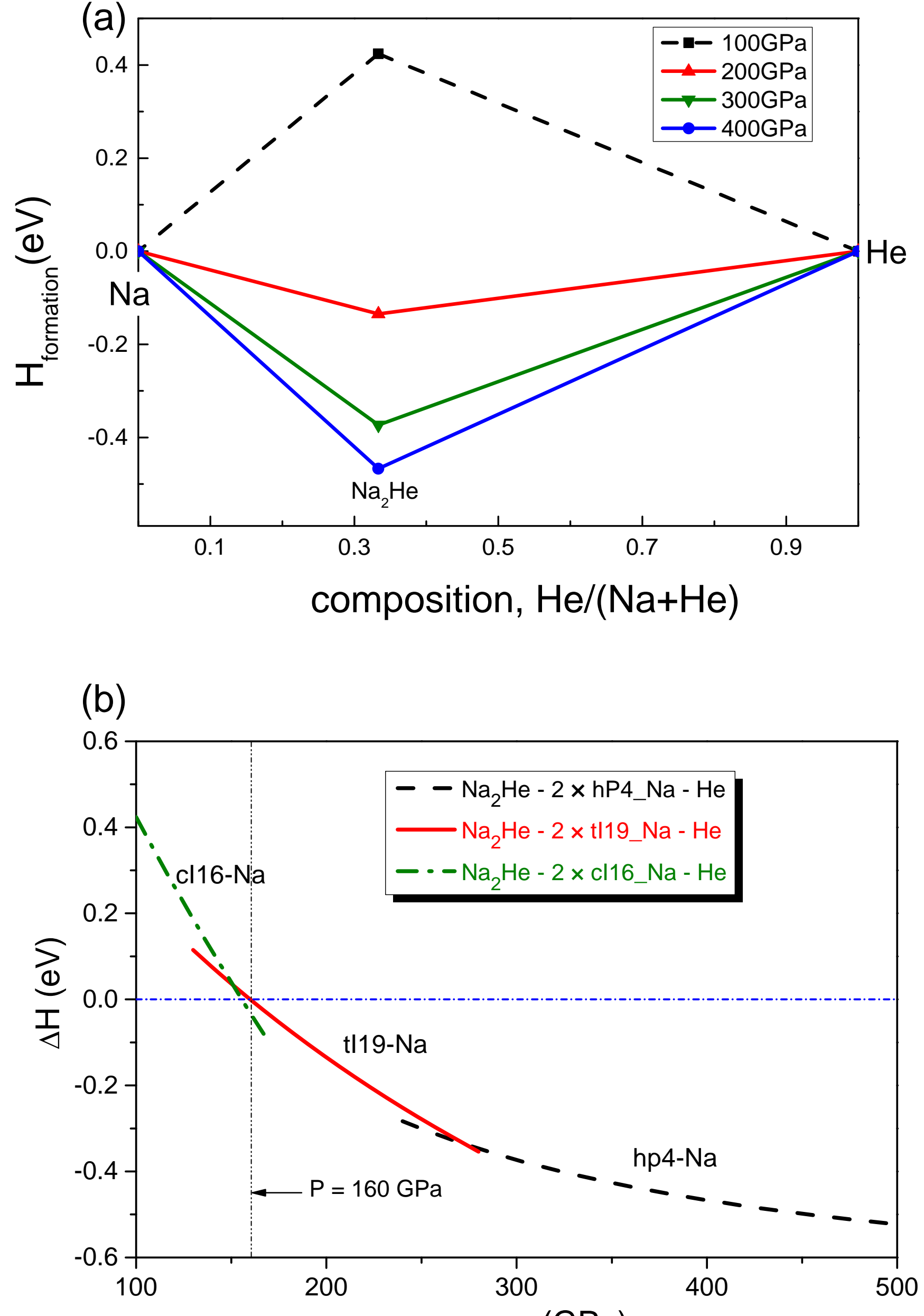

FIG. 1. Thermodynamics of the Na-He system. (a) Predicted convex hulls of the Na-He system, based on theoretical ground states of Na and He at each pressure (Refs. [13,17-19]). (b) Enthalpy of formation of $Na_2He$ as a function of pressure. Our calculated pressures of the cI16-tI19 and tI19-hP4 transitions of Na are 151 GPa and 273 GPa, respectively, similar to previous calculations [13].

$Na_2He$ becomes stable at a pressure near the cI16-tI19 transition of elemental Na (we also observed this in experiment, see below). The transition in Na has an underlying electronic stabilization mechanism related to the development of a pseudogap or gap at the Fermi energy (tI19 is a very poor metal, and hP4, the phase appearing on further increase of pressure, an insulator[13]). This hints at possible electronic stabilization of $Na_2He$, which has a surprisingly wide band gap in the entire pressure range of its stability. Insulating $Na_2He$ is predicted to form at a lower pressure than insulating hP4-Na: 160 GPa *vs* 273 GPa. The transition pressure for hP4-Na is significantly overestimated (the experimental value is 195 GPa[13]): DFT is known to underestimate stability fields of systems with localized electronic states, and one expects that $Na_2He$ will likewise become stable at pressure below the theoretical estimate of 160 GPa.

$Na_2He$ has only one ground-state structure (Fig. 2) in the whole pressure range explored here, i.e. from 160 to 1000 GPa. There are several equivalent ways to describe it: (i) Considering only positions of the atoms, this is a fluorite-type structure, a 3D-checkerboard with alternating He-filled and empty cubes formed by the Na atoms. However, fluorite structure is not dense and in all known compounds becomes unstable already at moderate pressures[20]. (ii) Just like hP4-Na[13], $Na_2He$ is an electride, i.e., ionic crystal with a strong Madelung field and interstitially localized electron (actually, an electron pair in both hP4-Na and $Na_2He$) playing the role of anion. If one also considers the positions of localized electron pairs (2e), this is a topologically very dense Heusler alloy structure ($AlCu_2Mn$-type, related to $Fe_3Al$-type): He atoms form a cubic close packing, in which all tetrahedral voids are filled by the Na atoms, and 2e fill all octahedral voids. Every He atom (and every 2e) is coordinated by eight Na atoms. Note that 2e form a cubic close packing of their own; in hP4-Na, they form a nearly perfect hexagonal close packing[13]. (iii) This structure can also be viewed as an ordered bcc superstructure formed by Na, He and 2e.

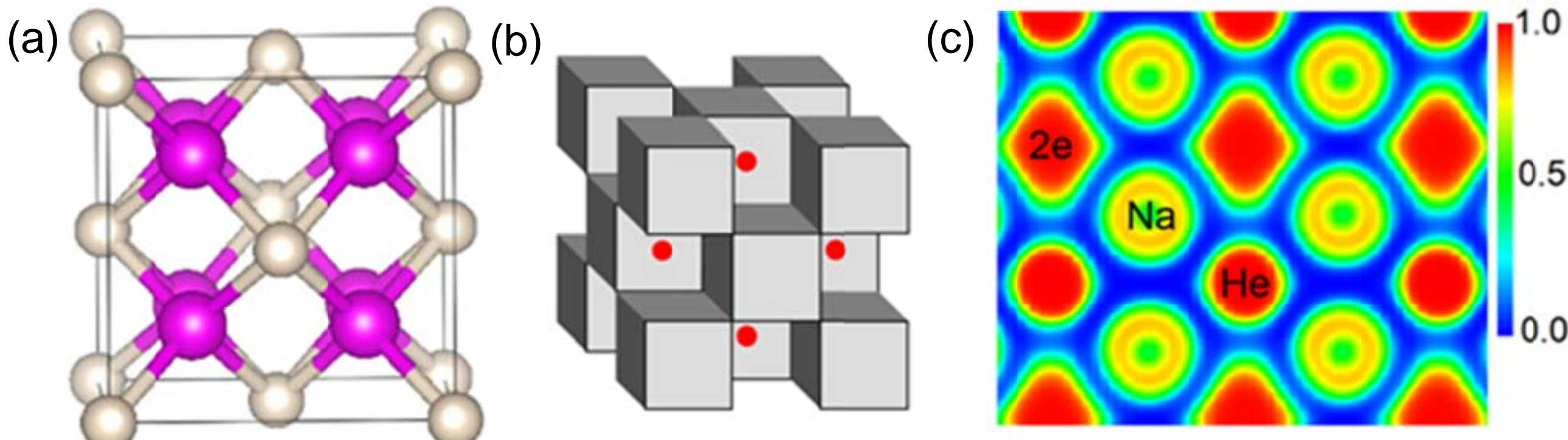

FIG. 2. Crystal structure of $Na_2He$ at 300 GPa. (a) ball-and-stick representation (pink and gray atoms represent Na and He, respectively) and (b) polyhedral representation, where half of $Na_8$ cubes are occupied by He atoms (shown by polyhedra), and half by 2e (shown by red spheres). (c) ELF plotted in the [110] plane at 300 GPa. This structure has space group *Fm-3m* with lattice parameter *a* = 3.95 Å at 300 GPa, and Na atoms occupying the Wyckoff position 8c (0.25,0.25,0.25) and He atoms occupying the 4*a* (0,0,0) positions.

In our experiments, Na was loaded in He medium in a laser-heated diamond anvil cell (DAC) and compressed up to 155 GPa (Supplementary Table S1). The sample was monitored using synchrotron X-ray diffraction (XRD), Raman spectroscopy, and visual observations. The latter verified (Fig. S3) that there was He in the DAC high-pressure cavity, as there were transparent colourless areas around Na sample, which remained to the highest pressure reached. Below ~110 GPa, only single crystal reflections of elemental Na were observed in XRD, and their positions agreed with previously reported structural data and equation of state (EOS) [18,21]. Above 113 GPa, we detected the appearance of new single-crystal reflections, which became stronger after laser heating to *T* >1500 K. Upon further compression, unreacted Na (which remains dominant) in the quenched sample showed the transition sequence cI16-oP8-tI19 typical of Na at *P* >113 GPa [18]. At *P* >140 GPa the transformation to tI19 phase was complete. Prolonged heating at 140 GPa yielded quasi-continuous diffraction lines (Fig. 3a,b), revealing substantial production of a new phase. Our laser heating experiments confirmed that compressed Na has low melting temperature, slightly above 300 K near 120 GPa[18,22]. Heating close to the melting temperature of He (~1500 K)[23] yields more reaction product, and yield increases during further heating.

New reflections were assigned to the predicted fluorite-type $Na_2He$ (Fig. 3a,b): they can be indexed in a cubic structure based on their positions and relative intensities (Table S2), with lattice parameter in good agreement with theory (Fig. 3c). Experiments show that $Na_2He$ has a much higher melting point than pure Na, perhaps above 1500 K at 140 GPa (Fig. 3a and 3b), where pure sodium melts slightly above 550 K [22,24], indicating very different energy landscapes and bonding types in Na and $Na_2He$. Experiments confirm stability of $Na_2He$, as it is denser than the mixture of Na and He (Fig. 3c) and crystallizes from sodium-helium melt. Our experiments traced this phase on decomposition down to 113 GPa, and its volume-pressure dependence is in good agreement with theoretical predictions (Fig. 3c). The increased yield of the new phase after laser heating (Fig. 3a,b) indicates that low yield of $Na_2He$ at near-room temperature is due to kinetic hindrance for the Na (fluid)-He (solid) reaction. Raman experiments on samples quenched to 300 K (Fig. S4) showed the presence of a new broad weak peak at 470 cm$^{-1}$ in addition to peaks which can be assigned to tI19-Na[22].

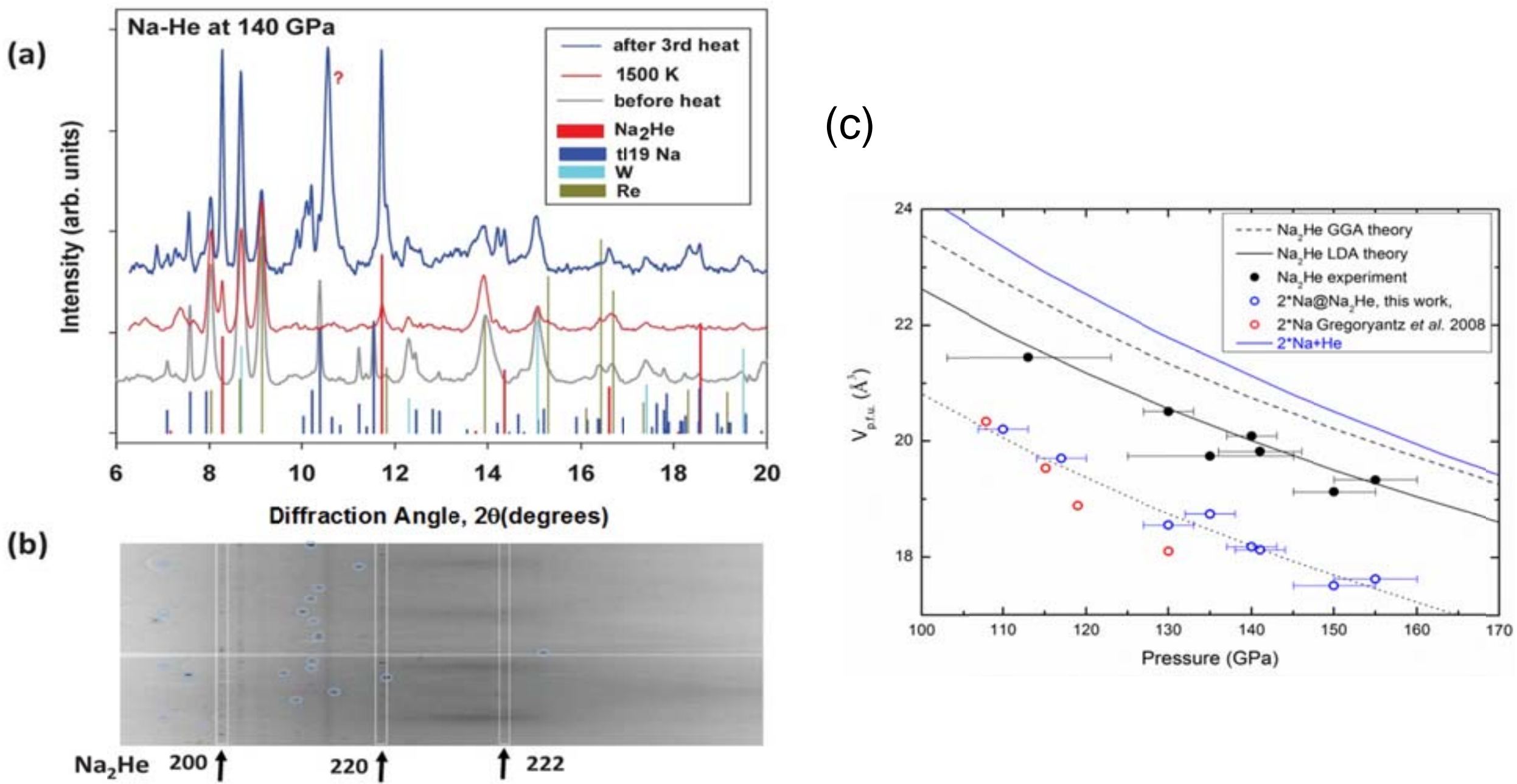

FIG. 3. Experimental data on $Na_2He$: (a,b) XRD at 140 GPa. (a) integrated XRD patterns before heating and after the second and third heating to approximately 1200 and >1500 K, respectively. Vertical ticks correspond to calculated XRD intensities of $Na_2He$, tI19 Na[18], Re (gasket) and W (pressure gauge). (b) 2D-image after the third heating in radial coordinates (cake), showing single-crystal reflections of tI19-Na and $Na_2He$, marked by light blue circles and white rectangles, respectively. This pattern was obtained during a continuous rotation of the DAC along the ω-axis to collect as many as possible single-crystal reflections of tI19-Na. Semi-continuous lines are from the rhenium gasket. Diffraction from $Na_2He$ and W consists of many small almost uniformly spaced spots. After the third heating a new almost continuous line appeared near 10.5°, which we assigned to yet unidentified reaction products. The X-ray wavelength is 0.31 Å. (c) Equation of state (EOS) of $Na_2He$ synthesized in a DAC at 113-150 GPa in comparison with the EOS of Na. Filled circles: experimental unit cell volumes of $Na_2He$. Open blue circles: volumes per 2 Na atom. Error bars correspond to the experimental pressure uncertainty due to pressure gradients and pressure measurements. Pressure was determined from XRD measurements of W marker in one experiment and using Raman of the stressed diamonds[25]. Solid blue line corresponds to a superposition of the EOS of Na and He (2Na+He) determined from experimental data[26]. Dashed and solid lines are the results of our GGA and LDA calculations, respectively. Dotted line: extrapolated EOS of fcc-Na from Ref.[21]. Open red circles: volumes of Na phases reported in Ref.[18].

Having experimentally confirmed the stability of this compound, we want to understand its origin. By analogy with other noble gases, one could expect helium to rather form stable compounds with the most electronegative atoms, or perhaps with molecular non-metallic elements such as H (e.g., Xe forms stable fluorides at ambient conditions, stable oxides under pressure[11], and compounds with H[27]).

However, due to the extremely high ionization potential of helium, the highest among all elements, helium does not react with oxygen or fluorine even at the extremely high pressure of 800 GPa. Xenon was predicted to react with electropositive elements (e.g. with Mg[12]) under pressure, but in contrast to Xe, He has zero electron affinity and is much less reactive.

It can be easily demonstrated that $Na_2He$ is not an inclusion compound. Formation of inclusion compounds involves little electronic redistribution, and host-guest interaction has only moderate effect on physical properties. $Na_2He$ is very different: its formation is strongly exothermic, and its Na-sublattice has a simple cubic structure with 1 valence electron per unit cell and must be metallic, but upon insertion of He a wide band gap opens up (Fig. 4). Electronic redistribution due to the insertion of He into the simple cubic sodium sublattice (Fig. 4) is also very large, again proving that this is not an inclusion compound. We observe a contraction of charge density towards all nuclei, but the main effect is its removal from the region occupied by helium and strong buildup in the empty $Na_8$ cubes. Strong non-nuclear charge density maxima allow us to call this compound an electride[28,29].

$Na_2He$ and hP4-Na differ fundamentally from the known low-pressure electrides [30], where interstitially localized electrons are unpaired and spin-polarized. Spin pairing increases the density, making electron-paired electrides a novel type of compounds stable under pressure. Valence and conduction bands are expected to broaden under pressure, leading to gap closure and pressure-induced metallization (Wilson model). However, both hP4-Na and $Na_2He$ display the opposite behavior: they are insulating throughout their stability fields, with band gaps increasing under pressure (Fig. 4b). With direct band gaps exceeding 1.8 eV at pressures above ~200 GPa, both $Na_2He$ and hP4-Na are expected to be optically transparent; for hP4-Na, this prediction was experimentally confirmed[13]. In our experiments, optical transparency of $Na_2He$ could not be verified because of the presence of unreacted Na shielding the transmitted light. Theoretically, $Na_2He$ has an even wider gap than hP4-Na at pressures below 230 GPa. Interstitial localization of valence electrons can be viewed as a result of overlap of core and valence orbitals of neighboring atoms, forcing valence electrons into the voids of the structure. At 300 GPa, the Na-Na distance is 1.98 Å, shorter than twice the core radius[31] (2.04 Å) of the Na atom. Compression leads to further localization of the interstitial electron pair, band narrowing and opening of the band gap[32]. Quantum mechanics predicts that interstitial electron pairs, once localized in space, will tend to adopt a spherical shape to minimize their kinetic energy, and our calculations indeed show nearly spherical electron localizations (Supplementary Fig. S5), characterized by volumes and radii, as normal atoms.

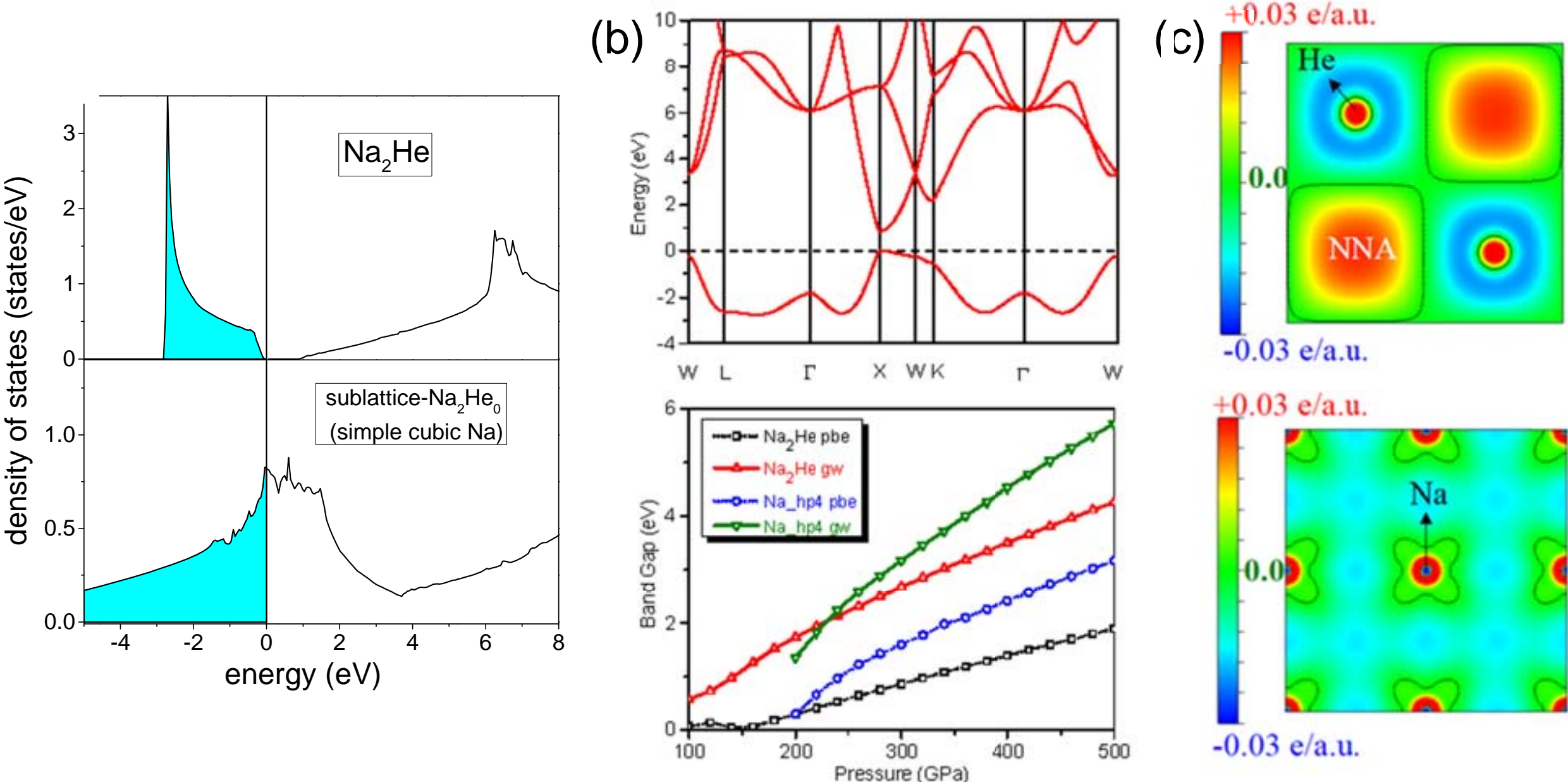

FIG. 4. Electronic structure of $Na_2He$. (a) Insertion of He opens a band gap (pure Na sublattice is metallic), (b) Band structure and band gap as a function of pressure. GW gaps are typically accurate to within 5-10% of experimental values[33]. (c) Electron density change induced by insertion of the He sublattice. (100) sections passing through He atoms (above) and Na atoms (below) are shown at 300 GPa. Acronym "NNA" (non-nuclear attractors) indicates interstitial electron localizations.

We can bring the electride description one step further, considering $Na_2He$ as a special kind of ionic crystal, stable because it satisfies Pauling's rules[34] for ionic structures. 1st Pauling's rule is satisfied here: ratios of radii (defined as the shortest distance to its Bader boundary[35,36]), $r_{He}/r_{Na}$~0.83 and $r_{2e}/r_{Na}$~0.75 are compatible with eightfold coordination (which requires ratios >0.732). Favorable atomic size ratio $r_{He}/r_{Na}$ alone is not sufficient to stabilize this compound: we find that although the $r_{Ne}/r_{K}$ ratio under pressure is closer to the ideal 0.732 for eightfold coordination (Supplementary Figs. S6 and S7 and Table S3), all K-Ne compounds are massively unstable. Taking into account positions and charges of "2e", one finds that $Na_2He$ also satisfies 2nd Pauling's rule (local electrostatic balance), while hP4-Na does not satisfy it - explaining why Na so readily reacts with He to form $Na_2He$ at high pressure.

Bader charges[35] (Table S5) of Na atoms in both hP4-Na and $Na_2He$ are close to +0.6, He atoms in $Na_2He$ have a small negative charge (~ -0.15*e*), while charges of "2e" are close to -1.1. For comparison, charge transfer in He@$H_2O$[9] , predicted at ultrahigh pressures, is just 0.03e. Further insight can be obtained by considering the change of properties of Bader atoms upon the formation of $Na_2He$ from the elements (Table 1). Below 300 GPa, all stabilization is due to volumetric gain (*PV*-term in the enthalpy); above 300 GPa energetic stabilization also comes into play. Table 1 shows large changes of the atomic energies as a result of electronic redistribution at 300 GPa. While He atoms expand and lower their energy, all the other atomic basins shrink, and their energy increases. The balance of different factors is quite complex, the greatest stabilizing factors being the decrease of the energy of the He atoms by 11.515 eV and shrinking of 2e (*PV* decrease by 0.966 eV).

TABLE 1. Change in atomic properties integrated over Bader basins for reaction (1) at 300 GPa –number of electrons *N*, atomic volume *V*, the *PV*-term, and the atomic energy *E*.

| | Total | Na1 | Na2 | He | 2e |
|---|---|---|---|---|---|
| $\Delta N$ | 0.000 | 0.007 | -0.067 | 0.151 | -0.094 |
| $\Delta V$ (Å$^3$) | -0.207 | -0.130 | -0.216 | 0.644 | -0.516 |
| $P\Delta V$ (eV) | -0.388 | -0.244 | -0.405 | 1.206 | -0.966 |
| $\Delta E$ (eV) | -0.018 | 3.090 | 7.764 | -11.515 | 0.642 |

We performed a simple computational experiment, taking a large fcc-supercell of He and replacing one He atom with Na as in Ref. [28,29]. We found that at pressures ~80 GPa, occupied Na-3s (HOMO) and empty He-2s (LUMO) energies become equal, enabling orbital mixing (Fig. S11) and some charge transfer from Na-3s to He-2s.

Further insight is provided by advanced methods, such as solid-state adaptive natural density partitioning (SSAdNDP)[37,38] and periodic natural bond orbital (NBO) methods[39,40] (see Table S6) and integrated crystal orbital Hamilton populations (ICOHP)[41,42] (Fig. S10). SSAdNDP and NBO give small, but puzzling, charge transfer from He-1s to Na-3p, and more expected transfers from He-1s to He-2s, and from Na-3s to Na-3p orbitals on increasing pressure. ICOHP shows that Na-Na and especially Na-2e (but not Na-He and He-2e) are the only important interactions, and they are both bonding direct covalent interactions. This conclusion is consistent with SSAdNDP and NBO analyses, where all electron density is assigned to the atoms, and electron pairs occupying empty $Na_8$ cubes emerge as eight-center two-electron (8c-2e) bonds. These bonds are formed by $sp^3$-hybridized atomic orbitals of sodium as derived from the NBO method[39,40]. If one considers only Na sublattice, on average there is just 1 electron located in every $Na_8$ cube. Upon insertion of He into half of all $Na_8$ cubes to form $Na_2He$, He atoms push electron density out of the filled $Na_8He$ cube to the neighboring empty $Na_8$ cube, enhancing the formation of 8c-2e bonds. Even at 0 GPa, He pushes out ~0.4 |e| from its cube, which helps to form 8c-2e bond with the occupation number (ON) of 1.40 |e| inside the neighboring empty cube. There are still about 0.6 |e|, which can be found in the $Na_8He$ cube. Upon increasing pressure, He continues to push out the remaining electron density into the empty $Na_8$ cube, increasing the ON of the 8c-2e bond to 1.80 |e| at 100 GPa, 1.88 |e| at 300 GPa and finally to 1.89 |e| at 500 GPa (Table S6). This 8-center 2-electron bonding is essential for stability of $Na_2He$. A complementary view of this is that $Na_2He$ is an ionic salt (electride), stable thanks to long-range electrostatic interactions, and crystal structure of which is in perfect harmony with the electronic redistribution.

Na is a light alkali metal, and is less reactive than the heavier K, Rb and Cs at ambient pressure. Yet, our calculations show the absence of thermodynamically stable K-He, Rb-He and Cs-He compounds at pressures below 1000 GPa. For Li, we do find that $Li_5He_2$ becomes stable at 780 GPa. The difference between light (Li and Na) and heavy (K, Rb and Cs) alkali metals was discussed by Winzenick *et al.*[43]: only heavy alkalis under pressure undergo an *s→d* electronic transition, making them "incipient transition metals". For example, the electronic configuration of K changes from $[Ar]4s^1$ to $[Ar]3d^1$, and the 3d-electron will be rather localized and able to penetrate the core, explaining reduced reactivity and absence of stable K-He compounds. Under pressure, Na paradoxically has lower electronegativity and higher reactivity than K. Indeed, we find that in A-He and A-Ne (A=Na, K) systems under pressure, Na-compounds have much lower enthalpies of formation (e.g., Supplementary Fig. S6). High reactivity of compressed Na is due to the interstitial electron pair and is consistent with what is known about electrides – that they have extremely low work functions and can be used as powerful reducing agents[44]. It is very interesting that, according to our calculations (and consistent with properties of known low-pressure electrides), the insulating electride phase of Na is more reactive than metallic Na; Li is only a weak electride, whereas heavier alkali metals (K, Rb, Cs) are not electrides and less reactive under pressure.

With this in mind, we hypothesized that $Na_2He$ with a bare interstitial electron pair might be stabilized by a strong acceptor of an electron pair – e.g., oxygen (see Ref. 45). Our structure searches showed that indeed $Na_2O$ has the same structure as hP4-Na and $Na_2HeO$ is isostructural with $Na_2He$, in both cases $O^{2-}$ (i.e. oxygen with the absorbed electron pair) occupies the position of "2e", and both can be considered as salts. Importantly, $Na_2HeO$ is thermodynamically stable in the Na-He-O system in the pressure range 15-106 GPa.

In conclusion, systematic search for stable compounds of helium has resulted in the prediction and experimental verification of a cubic phase $Na_2He$, stable from 113 GPa up to at least 1000 GPa. This phase is an electride, i.e. a crystal made of positively charged ionic cores and strongly localized valence electrons playing the role of anions. The insertion of He atoms pushes away the electron gas, leading to localization of valence electrons and formation of 8-center 2-electron bonds and opening of a wide band gap, just like for a salt-like compound. The predicted two compounds, $Na_2He$ and $Na_2HeO$, change the hitherto bare field of helium chemistry, provide new twists to the chemistry of noble gases, and will have impact on our understanding of chemical bonding, and of chemical processes that occur inside giant planets.

## METHODS

**Theory:** The evolutionary algorithm USPEX[16], used here for predicting new stable structures, searches for lowest-enthalpy structures at given pressure and is capable of predicting stable compounds and structures knowing just the chemical chemical elements involved. A number of applications[11,13,14,16,46] illustrate its power. Structure relaxations were performed using density functional theory (DFT) within the Perdew-Burke-Ernzerhof (PBE) functional[47] in the framework of the all-electron projector augmented wave (PAW) method[48] as implemented in the VASP code [49]. For Na atoms we used PAW potentials with 1.2 a.u. core radius and $2s^2 2p^6 3s$ electrons treated as valence; for He the core radius was 1.1 a.u. and $1s^2$ electrons were treated as valence. We used a plane-wave kinetic energy cutoff of 1000 eV, and the Brillouin zone was sampled with a resolution of $2\pi \times 0.06$ Å$^{-1}$, which showed excellent convergence of the energy differences, stress tensors and structural parameters. The first generation of structures was created randomly. All structures were relaxed at constant pressure and 0 K, and the enthalpy was used as fitness. The energetically worst structures (40%) were discarded and a new generation was created 30% randomly and 70% from the remaining structures through heredity, lattice mutation and transmutation.

To obtain atomic properties (Bader charges, atomic volumes and atomic energies), perform critical point analysis, compute Mulliken charges and deformation density maps we performed all-electron calculations using CRYSTAL14 code[50]. Triple-zeta quality Gaussian basis sets were used for all atoms, including also basis functions centered on non-nuclear charge density maxima positions, which are the centers of interstitial electron localizations (details of the basis set and grid used to sample direct and reciprocal space are reported in Supplementary Materials). The topological properties of charge density were obtained using the TOPOND code[51] incorporated into CRYSTAL14. Bader volumes and charges were also obtained using VASP and code from Ref. 36, and the results are essentially identical to those obtained using CRYSTAL14. Crystal Orbital Hamilton Population (COHP) analysis[41] was performed using periodic localized orbitals as implemented in the TB-LMTO-ASA framework[52]. We also explored the effects of temperature on stability using the quasiharmonic approximation, for which phonon calculations were performed for all promising structures using the PHONOPY code[53]; for each structure, phonons were computed at 20 different volumes to predict the Gibbs free energy.

**Experiment:** We loaded thin Na (3-5 μm) plates of 30 × 30 μm$^2$ dimensions in the DAC cavity of 30-40 μm diameter made in preindented to 20 μm thickness rhenium gasket in a glove box and then filled the rest of the cavity with He gas compressed to 1600 bars. Diamond anvils with 70-100 μm central tips beveled to 300 μm outside culet diameter were used. Synchrotron X-ray diffraction was monitored on pressure increase. Pressure was determined by measuring the position of the stressed first-order Raman diamond edge[54]. Laser heating was performed at above 120 GPa. X-ray diffraction patterns and radiometric temperature measurements were used to characterize the sample state *in situ*. Laser heating remains very local during this procedure as our radiometric measurements and finite element calculations show. Thus, we do not expect any reaction with a gasket material (which remains close to room temeperature during the heating) or with diamond anvils; this was verified by subsequent X-ray diffraction and Raman mapping of the sample cavity including areas near the gasket edge. Raman measurements were performed using 488, 532, and 660 nm lines of a solid-state laser. The laser probing spot dimension was 4 μm. Raman spectra were analyzed with a spectral resolution of 4 $cm^{-1}$ using a single-stage grating spectrograph equipped with a CCD array detector. X-ray diffraction was measured in a double-sided laser heating system at the undulator XRD beamline at GeoSoilEnviroCARS, APS, Chicago and Extreme Conditions Beamline P02.2 at DESY (Germany), which have online laser heating capabilities. Temperature was determined spectroradiometrically. The X-ray probing beam size was ~2-5 μm in both beamlines.

artem.oganov@sunysb.edu

xfzhou@nankai.edu.cn

htwang@nankai.edu.cn/htwang@nju.edu.cn

1 Stevenson D. J. Metallic helium in massive planets. *Proc. Natl. Acad. Sci.* **105**, 11035-11036 (2008).

2 Huheey J. E., Keiter E. A., Keiter R. L. & Medhi O. K. *Inorganic chemistry: principles of structure and reactivity*. (Harper & Row New York, 1983).

3 Hotop H. & Lineberger W. C. Binding Energies in Atomic Negative Ions: II. *J. Phys. Chem. Ref. Data* **14**, 731-750 (1985).

4 Hiby J. W. Massenspektrographische Untersuchungen an Wasserstoff- und Heliumkanalstrahlen (H3+, H2-, HeH+, HeD+, He-). *Annalen der Physik* **426**, 473-487 (1939).

5 Wong M. W. Prediction of a metastable helium compound: HHeF. *J. Am. Chem. Soc* **122**, 6289-6290 (2000).

6 Grochala W. On chemical bonding between helium and oxygen. *Pol. J. Chem.* **83**, 87-122 (2009).

7 Tariq N., Taisan N., Singh V. & Weinstein J. D. Spectroscopic Detection of the LiHe Molecule. *Phys. Rev. Lett.* **110**, 153201 (2013).

8 Loubeyre P., Jean-Louis M., LeToullec R. & Charon-Gerard L. High pressure measurements of the He-Ne binary phase diagram at 296 K: Evidence for the stability of a stoichiometric Ne(He)2 solid. *Phys. Rev. Lett.* **70**, 178-181 (1993).

9 Liu H., Yao Y. & Klug D. D. Stable structures of He and H2O at high pressure. *Phys. Rev. B* **91**, 014102 (2015).

10 Hermann A. & Schwerdtfeger P. Xenon Suboxides Stable under Pressure. *J. Phys. Chem. Lett.* **5**, 4336-4342 (2014).

11 Zhu Q. *et al.* Stability of xenon oxides at high pressures. *Nat. Chem.* **5**, 61-65 (2012).

12 Miao M.-s. *et al.* Anionic Chemistry of Noble Gases: Formation of Mg–NG (NG = Xe, Kr, Ar) Compounds under Pressure. *J. Am. Chem. Soc.* **137**, 14122-14128 (2015).

13 Ma Y. *et al.* Transparent dense sodium. *Nature* **458**, 182-185 (2009).

14 Zhang W. *et al.* Unexpected Stable Stoichiometries of Sodium Chlorides. *Science* **342**, 1502-1505 (2013).

15 Lyakhov A. O., Oganov A. R. & Valle M. Crystal structure prediction using evolutionary approach. *Modern Methods of Crystal Structure Prediction(ed. A.R. Oganov)*, 147-180 (2010).

16 Oganov A. R. & Glass C. W. Crystal structure prediction using ab initio evolutionary techniques: Principles and applications. *J. Chem. Phys.* **124**, 244704-244704 (2006).

17 Gregoryanz E., Degtyareva O., Somayazulu M., Hemley R. J. & Mao H.-k. Melting of dense sodium. *Phys. Rev. Lett.* **94**, 185502 (2005).

18 Gregoryanz E. *et al.* Structural diversity of sodium. *Science* **320**, 1054-1057 (2008).

19 McMahon J. M., Morales M. A., Pierleoni C. & Ceperley D. M. The properties of hydrogen and helium under extreme conditions. *Rev. Mod. Phys.* **84**, 1607 (2012).

20 Gerward L. *et al.* X-ray diffraction investigations of CaF2 at high pressure. *J. Appl. Crystallogr.* **25**, 578-581 (1992).

21 Hanfland M., Loa I. & Syassen K. Sodium under pressure: bcc to fcc structural transition and pressure-volume relation to 100 GPa. *Phys. Rev. B* **65**, 184109 (2002).

22 Marqués M. *et al.* Optical and electronic properties of dense sodium. *Phys. Rev. B* **83**, 184106 (2011).

23 Santamaría-Pérez D., Mukherjee G. D., Schwager B. & Boehler R. High-pressure melting curve of helium and neon: Deviations from corresponding states theory. *Phys. Rev. B* **81**, 214101 (2010).

24 Gregoryanz E., Degtyareva O., Somayazulu M., Hemley R. J. & Mao H.-k. Melting of Dense Sodium. *Phys. Rev. Lett.* **94**, 185502 (2005).

25 Akahama Y. & Kawamura H. Pressure calibration of diamond anvil Raman gauge to 310GPa. *J. Appl. Phys.* **100**, 043516 (2006).

26 Loubeyre P. *et al.* Equation of state and phase diagram of solid 4He from single-crystal x-ray diffraction over a large P-T domain. *Phys. Rev. Lett.* **71**, 2272-2275 (1993).

27 Somayazulu M. *et al.* Pressure-induced bonding and compound formation in xenon–hydrogen solids. *Nat. Chem.* **2**, 50-53 (2010).

28 Miao M.-S. & Hoffmann R. High Pressure Electrides: A Predictive Chemical and Physical Theory. *Acc. Chem. Res.* **47**, 1311-1317 (2014).

29 Miao M.-s. & Hoffmann R. High-Pressure Electrides: The Chemical Nature of Interstitial Quasiatoms. *J. Am. Chem. Soc.* **137**, 3631-3637 (2015).

30 Dye J. L. Electrons as anions. *Science* **301**, 607-608 (2003).

31 Shannon R. t. & Prewitt C. T. Effective ionic radii in oxides and fluorides. *Acta Cryst. B* **25**, 925-946 (1969).

32 Rousseau B. & Ashcroft N. W. Interstitial Electronic Localization. *Phys. Rev. Lett.* **101**, 046407 (2008).

33 Shishkin M. & Kresse G. Self-consistent GW calculations for semiconductors and insulators. *Phys. Rev. B* **75**, 235102 (2007).

34 Pauling L. The principles determining the structure of complex ionic crystals. *J. Am. Chem. Soc* **51**, 1010-1026 (1929).

35 Bader R. F. *Atoms in molecules*. (Wiley Online Library, 1990).

36 Henkelman G., Arnaldsson A. & Jonsson H. A fast and robust algorithm for Bader decomposition of charge density. *Comput. Mat. Sci.* **36**, 354-360 (2006).

37 Galeev T. R., Dunnington B. D., Schmidt J. & Boldyrev A. I. Solid state adaptive natural density partitioning: a tool for deciphering multi-center bonding in periodic systems. *Phys. Chem. Chem. Phys.* **15**, 5022-5029 (2013).

38 Zubarev D. Y. & Boldyrev A. I. Developing paradigms of chemical bonding: adaptive natural density partitioning. *Phys. Chem. Chem. Phys.* **10**, 5207-5217 (2008).

39 Dunnington B. D. & Schmidt J. Generalization of Natural Bond Orbital Analysis to Periodic Systems: Applications to Solids and Surfaces via Plane-Wave Density Functional Theory. *J. Chem. Theory Comput.* **8**, 1902-1911 (2012).

40 Foster J. & Weinhold F. Natural hybrid orbitals. *J. Am. Chem. Soc.* **102**, 7211-7218 (1980).

41 Dronskowski R. & Bloechl P. E. Crystal orbital Hamilton populations (COHP): energy-resolved visualization of chemical bonding in solids based on density-functional calculations. *J. Phys. Chem.* **97**, 8617-8624 (1993).

42 Andersen O. K. & Jepsen O. Explicit, First-Principles Tight-Binding Theory. *Phys. Rev. Lett.* **53**, 2571-2574 (1984).

43 Winzenick M., Vijayakumar V. & Holzapfel W. B. High-pressure x-ray diffraction on potassium and rubidium up to 50 GPa. *Phys. Rev. B* **50**, 12381-12385 (1994).

44 Dye J. L. Electrides: early examples of quantum confinement. *Acc. Chem. Res.* **42**, 1564-1572 (2009).

45 Vegas Á. & Mattesini M. Towards a generalized vision of oxides: disclosing the role of cations and anions in determining unit-cell dimensions. *Acta Crystallogr. Sect. B: Struct. Sci.* **66**, 338-344 (2010).

46 Oganov A. R. *et al.* Ionic high-pressure form of elemental boron. *Nature* **457**, 863-867 (2009).

47 Perdew J. P., Burke K. & Ernzerhof M. Generalized gradient approximation made simple. *Phys. Rev. Lett.* **77**, 3865-3868 (1996).

48 Blochl P. E. Projector augmented-wave method. *Phys. Rev. B* **50**, 17953 (1994).

49 Kresse G. & Furthmuller J., rgen. Efficiency of ab-initio total energy calculations for metals and semiconductors using a plane-wave basis set. *Comput. Mat. Sci.* **6**, 15-50 (1996).

50 Dovesi R. *et al. CRYSTAL14 User's Manual*. (University of Torino, 2014).

51 Gatti C., Saunders V. R. & Roetti C. Crystal field effects on the topological properties of the electron density in molecular crystals: The case of urea. *J. Chem. Phys.* **101**, 10686-10696 (1994).

52 Krier G., Jepsen O., Burkhardt A. & Andersen O. *The TB-LMTO-ASA Program.* (1995).
53 Togo A., Oba F. & Tanaka I. First-principles calculations of the ferroelastic transition between rutile-type and CaCl2-type SiO2 at high pressures. *Phys. Rev. B* **78**, 134106 (2008).
54 Akahama Y. & Kawamura H. High-pressure Raman spectroscopy of diamond anvils to 250GPa: method for pressure determination in the multimegabar pressure range. *J. Appl. Phys.* **96**, 3748-3751 (2004).

## ACKNOWLEDGEMENTS

This work was supported by the China Scholarship Council (grant 201206200030), National Science Foundation (grant EAR-1114313), DARPA (grant W31P4Q1210008), Special Program for Applied Research on Super Computation of the NSFC-Guangdong Joint Fund (the second phase), Russian Science Foundation (grant 16-13-10459), National 973 Program of China (grant 2012CB921900), and Foreign Talents Introduction and Academic Exchange Program (grant B08040). X.F.Z acknowledges funding from the National Science Foundation of China (grant 11174152), and the Program for New Century Excellent Talents in University (grant NCET-12-0278). Calculations were performed at the Tianhe II supercomputer in Guangzhou and supercomputer of the Center for Functional Nanomaterials, Brookhaven National Laboratory, which is supported by the U.S. Department of Energy, Office of Basic Energy Sciences, under contract No. DE-AC02-98CH10086. GeoSoilEnviroCARS is supported by the National Science Foundation - Earth Sciences (EAR- 1128799) and Department of Energy - Geosciences (DE-FG02-94ER14466). Use of the Advanced Photon Source was supported by the U. S. Department of Energy, Office of Science, Office of Basic Energy Sciences, under Contract No. DE-AC02-06CH11357. PETRA III at DESY is a member of the Helmholtz Association (HGF). The research leading to these results has received funding from the European Community's Seventh Framework Programme (FP7/2007- 2013) under grant agreement n° 312284. Work of E.S. was performed under the auspices of the U. S. Department of Energy by Lawrence Livermore National Security, LLC under Contract DE-AC52-07NA27344. A.F.G. acknowledges support from the National Natural Science Foundation of China (grant 21473211), the Chinese Academy of Sciences (grant YZ201524), and the Chinese Academy of Sciences visiting professorship for senior international scientists (grant 2011T2J20) and Recruitment Program of Foreign Experts. S.L. was partly supported by state assignment project No. 0330-2014-0013.

## AUTHOR CONTRIBUTIONS

X.D. and A.R.O. designed research. X.D., G.S. and I. P. performed and analyzed the calculations, V.L.D. and R.D. carried out COHP analyses, A.G. designed experiments. S.L. and A.G. loaded the sample, A.G., E.S., S.L., V.P., and Z.K. performed the experiment, E.S. and A.G. analyzed the experimental data. G.R.Q., Q.Z., X.F.Z. and A. B. assisted with calculations. All authors contributed to interpretation and discussion of the data. X.D., A.R.O, A.G., G.S., I. P., A. B. and H.T.W wrote the manuscript.